# Microwave photonic short-time Fourier transform based on stabilized period-one nonlinear laser dynamics and stimulated Brillouin scattering

Sunan Zhang[a], Taixia Shi[a], Lizhong Jiang[b], Yang Chen[a,*]

[a] Shanghai Key Laboratory of Multidimensional Information Processing, School of Communication and Electronic Engineering, East China Normal University, Shanghai 200241, China
[b] Shanghai Radio Equipment Research Institute, Shanghai 201109, China
*Correspondence to: Y. Chen, ychen@ce.ecnu.edu.cn.

**ABSTRACT**
A microwave photonic short-time Fourier transform (STFT) system based on stabilized period-one (P1) nonlinear laser dynamics and stimulated Brillouin scattering (SBS) is proposed. By using an optoelectronic feedback loop, the frequency-sweep optical signal generated by the P1 nonlinear laser dynamics is stabilized, which is further used in conjunction with an optical bandpass filter implemented by stimulated Brillouin scattering (SBS) to achieve the frequency-to-time mapping of microwave signals and the final STFT. By comparing the experimental results with and without optoelectronic feedback, it is found that the time-frequency diagram of the signal under test (SUT) obtained by STFT is clearer and more regular, and the frequency of the SUT measured in each frequency-sweep period is more accurate. The mean absolute error is reduced by 50% under the optimal filter bandwidth.

## 1. Introduction

Spectrum sensing holds a key position across diverse domains, encompassing wireless communication, electronic warfare, and intelligent transportation systems [1-3]. Taking electronic warfare as an example, spectrum sensing is used to perceive and analyze the surrounding spectrum environment, so as to promptly detect surrounding electromagnetic signals, such as radar signals, electromagnetic interferences, etc. Generally speaking, spectrum sensing refers to the measurement of signal spectrum. Some existing methods, such as short-time Fourier transform (STFT), wavelet transform, etc., for acquiring time-frequency information can also be used for spectrum sensing. However, because these methods are generally implemented in the digital domain, the analysis of large-bandwidth microwave signals faces problems such as limited sampling rate and excessive data volume.

    To solve the above problems, in recent years, the analog time-frequency analysis of microwave signals, especially STFT, based on microwave photonics has gained widespread attention, which has two main technical routes: 1) Frequency-to-time mapping (FTTM) based on dispersion medium [4-7]; 2) FTTM based on high-speed optical frequency-sweep and filtering [8-11]. Earlier dispersion-based schemes [4,

5] achieved very good time resolution, but their analysis bandwidth was limited by the amount of dispersion and only reached 2.43 GHz. In [6], by using a complex multi-level waveform with a step size of approximately 22 ps and a bandwidth exceeding 46 GHz, the instantaneous analysis bandwidth was extended to tens of GHz. However, such a complex waveform is difficult to generate in practical applications. In [7], a fiber loop with a length of 0.38 meters was used, which provided an equivalent high dispersion of about $4.5\times10^5$ ps/nm. Although the length of the dispersion medium is greatly reduced, the instantaneous analysis bandwidth is only 530 MHz. Besides, the reconfigurability of most dispersion-based time-frequency analysis methods is severely limited by the dispersion medium.

Compared to the dispersion-based methods, the methods via FTTM based on high-speed optical frequency-sweep and filtering [8-11] have a significant advantage in the instantaneous analysis bandwidth, as well as the reconfigurability. The first corresponding STFT scheme was given in [8], in which the high-speed optical frequency-sweep signal was generated using an electrical frequency-sweep signal, and the filtering was implemented in the optical domain using the stimulated Brillouin scattering (SBS) gain spectrum. An instantaneous analysis bandwidth of 12 GHz and a frequency resolution of around 60 MHz was realized. In [9], filtering in the optical domain was replaced by an electrical bandpass filter. Nevertheless, the high-speed frequency-sweep optical signal in [8, 9] is achieved by using a high-cost electrical frequency-sweep signal, which also makes it difficult to apply to real systems. Therefore, in [10, 11], we proposed two schemes to realize the analog STFT of microwave signals through all-optical frequency-sweeping. Nevertheless, the method based on directly modulating a distributed feedback (DFB) laser in [10] is limited in analysis bandwidth. The method based on optical injection in [11] lacks stability in the generated signals, resulting in poor quality of the time-frequency diagram obtained from time-frequency analysis and large measurement error from the frequency measurement. Furthermore, although the scheme in [11] replaced the SBS filtering in the previous scheme with phase-shifter fiber Bragg grating (PS-FBG) filtering, making the structure simpler, from the perspective of flexibility, the scheme using SBS filtering can adjust the filter bandwidth to an appropriate value based on the frequency-sweep rate of the optical frequency-sweep signal to achieve the best analysis performance.

In this letter, a microwave photonic STFT system based on stabilized period-one (P1) nonlinear laser dynamics [12] and SBS filtering is proposed. The frequency-sweep optical signal generated by the P1 nonlinear laser dynamics is stabilized using an optoelectronic feedback loop while the nonlinearity of the signal is post-compensated. Meanwhile, the filtering is implemented by a frequency-sweep SBS pump signal to achieve tunable filter bandwidth. It is found that by using the optoelectronic feedback loop, the time-frequency diagram of the signal under test (SUT) obtained by STFT is clearer and more regular, and the frequency of SUT measured in each frequency-sweep period is more accurate. The mean absolute error (MAE) is reduced by 50% under the optimal filter bandwidth.

## 2. Principle and experimental results

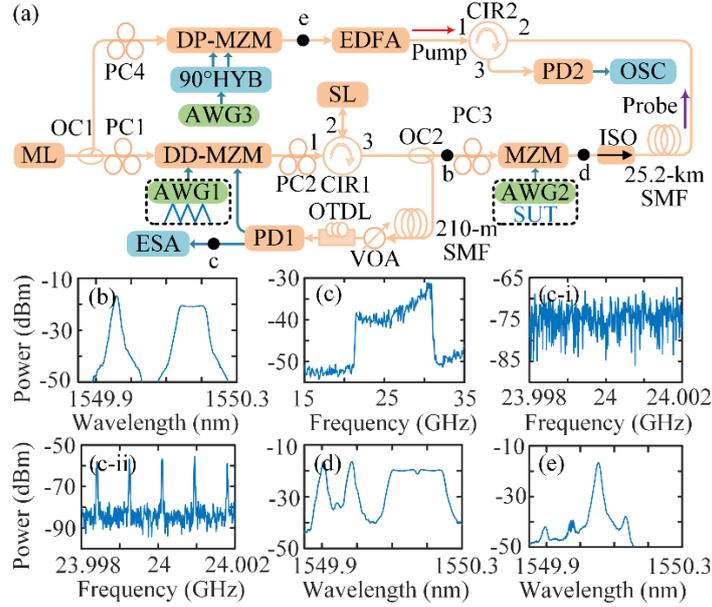

Fig. 1 (a) Schematic diagram of the proposed system. (b)-(e) Spectra at corresponding locations marked in (a). Zoomed-in-view of the electrical spectra of the frequency-sweep signal (c-i) without and (c-ii) with round-trip time matching at location c.

The schematic of the system is shown in Fig. 1(a). A laser diode (LD, ID Photonics CoBriteDX1-1-C-H01-FA) with a wavelength of 1549.973 nm and an output power of 15.5 dBm is used as the master laser (ML). Its output is divided into two paths via an optical coupler (OC1, 10:90). In the lower probe branch, one output of OC1 is injected into a dual-drive Mach–Zehnder modulator (DD-MZM, Fujitsu FTM7937EZ200) through a polarization controller (PC1). A triangular-shaped control signal from a low-speed arbitrary waveform generator (AWG1, RIGOL DG2052) is sent to one arm of the DD-MZM to control the optical power injected into the slave laser (SL), via an optical circulator (CIR1). By properly controlling the injection strength, a broadband nonlinear frequency-sweep optical signal is generated via the P1 nonlinear laser dynamics, as shown in Fig. 1(b). The output power of SL is divided into two parts by OC2. One output from OC2 is fed back to DD-MZM through a 210-m single-mode fiber (SMF), a variable optical attenuator (VOA), an optical tunable delay line (OTDL), and a photodetector (PD1, u2t MPRV1331A). Here, the VOA is used to optimize the total feedback strength. By tuning the OTDL and adjusting the period $T$ of the waveform from AWG1, the Fourier domain mode locking (FDML) [13] condition of the optoelectronic feedback can be guaranteed. When the FDML condition is satisfied, the performance of the optical frequency-sweep signal can be greatly improved. As can be seen from Fig. 1(c), a nonlinear frequency-sweep electrical signal is generated from PD1. When the round-trip time $\tau$ of the optoelectronic feedback loop fails to align with $T$, no comb-like patterns are visible in the electrical spectrum depicted in Fig. 1(c-i). However, upon achieving a precise match between $\tau$ and $T$, a distinct comb-like spectrum exhibiting a high signal-to-noise ratio (SNR) is obtained, as evident in Fig. 1(c-ii), which means a stable nonlinear frequency-sweep optical signal is generated at the output port of OC2. It should be noted that here we use a triangular wave as the driving signal instead of a sawtooth wave to avoid the effect of the rapid transition of sawtooth waveform on the generation of the frequency-sweep optical signal. Therefore, the nonlinear frequency-sweep optical signal sweeps from low-frequency $f_{s1}$ to high-frequency $f_{s2}$ within the time interval from 0 to $T/2$, and then sweeps back from $f_{s2}$ to $f_{s1}$ within the time interval of $T/2$ to $T$. We only use half of each period to implement the STFT and frequency measurement functions.

The other output from OC2 passes through PC3 and is then injected into an MZM (Fujitsu FTM7938EZ), in which the SUT is carrier-suppressed double-sideband (CS-DSB) modulated by the SUT generated from AWG2 (Keysight M8190A). The spectrum of the optical signal from the MZM when the SUT is a 5-GHz single-tone signal is shown in Fig. 1(d). The output of the MZM is injected into a 25.2-km SMF through an optical isolator (ISO) and used as the probe wave. The SMF is used as a nonlinear medium for SBS interaction, in which the probe wave interacts with the counter-propagating pump wave from the upper pump branch.

In the upper pump path, the optical carrier from OC1 is carrier-suppressed lower single-sideband modulated at a dual-parallel Mach–Zehnder modulator (DP-MZM, Fujitsu FTM7961EX) by an electrical frequency-sweep signal from AWG3 (Keysight M8195A) to generate an optical frequency-sweep pump wave, as shown in Fig. 1(e), which is sent to the 25.2-km SMF via an erbium-doped fiber amplifier (Amonics, EDFA-PA-35-B) and CIR2. The SBS gain spectrum functions as an optical filter to implement the FTTM of the system. By adjusting the frequency-sweep range of the pump wave, the SBS gain bandwidth can be effectively controlled. Then, the optical signal after SBS interaction and FTTM is detected in PD2 (Nortel PP-10G) to convert the optical pulses to electrical pulses, which are monitored by an oscilloscope (OSC, R&S RTO2032). By combining the frequency information obtained from the electrical pulses in multiple periods from 0 to $T/2$, the time-frequency diagram of SUT is obtained.

In this system, the SBS gain spectrum controlled by AWG3 should be properly positioned so that different frequency analysis ranges can be achieved. However, the instantaneous analysis bandwidth is only determined by the nonlinear frequency-sweep optical signal bandwidth ($f_{s2} - f_{s1}$) in the probe path. If the SBS gain spectrum is on the right side of $f_{s2}$, i.e., $f_0 - f_{BFS} > f_{s2}$, where $f_0$ is the center frequency of the frequency-sweep pump wave and $f_{BFS}$ is the Brillouin frequency shift, the frequency analysis range is from $f_0 - f_{BFS} - f_{s2}$ to $f_0 - f_{BFS} - f_{s1}$.

In the experiment, a triangular wave with a peak-to-peak amplitude of 3.6 V and a repetition rate of 847.45 kHz is employed, so a frequency-sweep optical signal with a period of 1.18-$\mu$s is generated. The repetition rate of the triangular wave is carefully selected according to the total length of the optoelectronic feedback loop. Under these circumstances, a stable comb-like spectrum can be observed via an electrical spectrum analyzer (ESA, R&S, FSP-40). The sweep bandwidth of the nonlinear frequency-sweep optical signal is around 8.4 GHz. By tuning the frequency relationship between $f_0$ and $f_{s1}$, the proposed STFT system can analyze signals from 3 to 11.4 GHz. Since the average chirp rate of the nonlinear frequency-sweep optical signal is 14.23 GHz/$\mu$s, in the pump path, the pump wave is configured with a sweep bandwidth of 150 MHz [14] to achieve a better performance. The SUT is a dual-chirp linearly frequency-modulated (LFM) signal. This signal is produced by mixing a 200-$\mu$s single-chirp LFM signal, ranging from 0 to 3 GHz, with a 6-GHz local oscillator. It is not directly generated from AWG2 because AWG2 can only generate a signal within 4 GHz. The time-frequency diagram of the SUT obtained by the STFT system is shown in Fig. 2, in which the white dotted line represents the theoretical position of the SUT time-frequency relationship. When the feedback loop is disconnected and the nonlinearity is not compensated, as shown in Fig. 2(a), the obtained time-frequency diagram deviates from the theoretical one due to the nonlinearity of the frequency-sweep optical signal generated from P1 nonlinear laser dynamics.

To address the distortion issue in STFT results caused by the nonlinear frequency-sweep optical signal,

we have employed a post-compensation technique. By inputting a known LFM signal to the STFT system and recording the resulting distorted time-frequency diagram, a one-to-one correspondence between the input frequency and output frequency under the action of the nonlinear frequency-sweep optical signal can be derived. Therefore, when measuring any unknown SUT, the distorted time-frequency diagram can be corrected using this correspondence to obtain the correct time-frequency curve.

As can be seen from Fig. 2(b), the time-frequency diagram after post-compensation is in complete agreement with the theoretical one, which proves the feasibility of the post-compensation method. Due to the large frequency range of the vertical coordinate in Fig. 2, the improvement brought by the optoelectronic feedback is not observed in the figure. Therefore, in the following experiments, we will only analyze signals within a 4-GHz bandwidth to better observe the performance improvement brought by the feedback. Moreover, except for Fig. 2(a), all the time-frequency diagrams shown in the rest of this letter are the results after post-compensation.

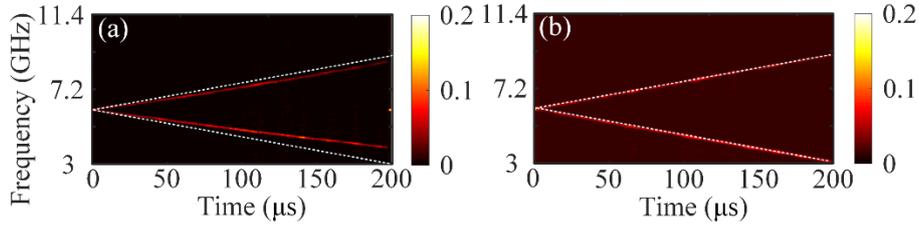

Fig. 2 Measured time-frequency diagrams of the dual-chirp LFM signal ranging from 3 GHz to 9 GHz (a) without and (b) with feedback and compensation.

In the results above, the SBS gain spectrum is broadened by using a 150-MHz sweep pump wave, so a very good frequency resolution can be observed in Fig. 2. Here the frequency-sweep range of the pump wave is set to 90, 150, and 270 MHz, respectively. The instantaneous analysis bandwidth of the system is changed to 10 GHz, while an LFM signal with a duration of 200 μs and a bandwidth of 4 GHz is chosen as the SUT. In this case, the average chirp rate of the nonlinear frequency-sweep optical signal is 16.95 GHz/$\mu$s. As given in [14], when the sweep chirp rate of the frequency-sweep optical signal is over 10 GHz, the best frequency resolution will be getting closer and closer. Because the SBS gain bandwidth range that can approach the best frequency resolution is tens of MHz, we take the 150MHz bandwidth as an example here. Under this bandwidth, both the chirp rate of 14.23 GHz/$\mu$s mentioned earlier and the chirp rate of 16.95 GHz/$\mu$s in this experiment can achieve a good resolution. As can be observed from Fig. 3, the time-frequency diagrams obtained when the sweep range of the pump wave is 150 MHz have significantly better frequency resolution compared to the cases when the sweep range of the pump wave is 90 MHz and 270 MHz. Besides, by comparing the results without and with the optoelectronic feedback, it can be seen that after adopting the feedback loop, the time-frequency diagram becomes significantly smoother and more regular. This strongly demonstrates the improvement of the optoelectronic feedback in enhancing the stability of the generated optical frequency-sweep signal and improving the time-frequency analysis results.

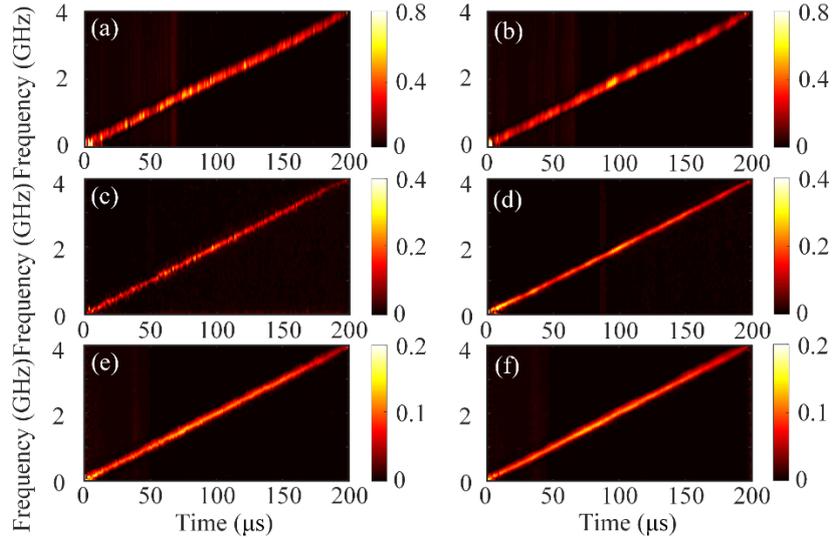

Fig. 3 Measured time-frequency diagrams of an LFM signal with a bandwidth ranging from 0 to 4 GHz (a), (c), (e) without and (b), (d), (f) with feedback. The frequency-sweep range of the pump wave is (a), (b) 90 MHz, (c), (d) 150 MHz, (e), (f) 270 MHz.

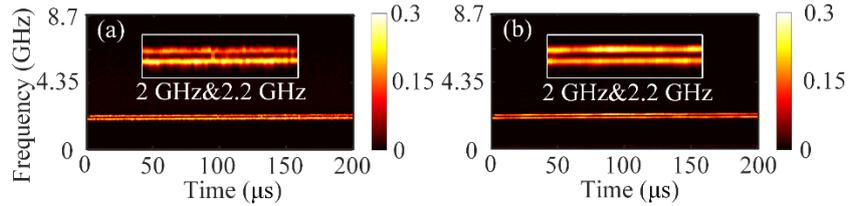

Fig. 4 Measured time-frequency diagrams of a two-tone signal with frequencies of 2.0 GHz and 2.2 GHz (a) without and (b) with feedback.

The frequency resolution is further demonstrated using a two-tone test. The instantaneous analysis bandwidth of the system is 8.7 GHz. The frequency-sweep range of the pump wave is also 150 MHz. Fig. 4 shows the measured time-frequency diagrams of the two-tone signal with a frequency interval of 200 MHz. It is observed that the frequency resolution is significantly below 200 MHz under a sweep chirp rate of around 14.75 GHz/μs and it is better when the optoelectronic feedback is incorporated.

Then, multi-format signals are analyzed by the proposed system and the frequency-sweep range of the pump wave is also 150 MHz. The obtained time-frequency diagrams of a dual-chirp LFM signal, a nonlinearly frequency-modulated signal, a step-frequency signal, and a frequency-hopping signal are shown in Fig. 5. As can be seen, the time-frequency information of different signal formats is well recovered, and the results with optoelectronic feedback shows significant better and clearer results.

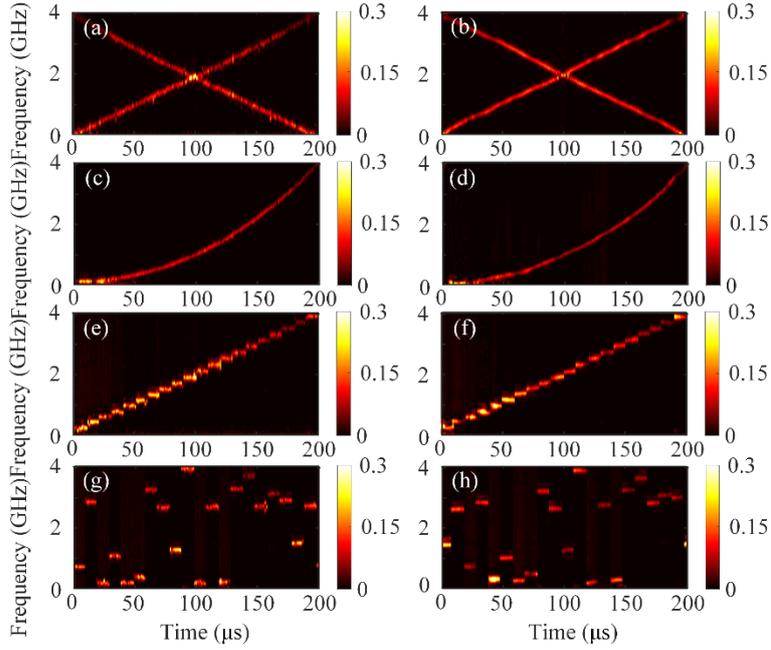

Fig. 5 Measured time-frequency diagrams of (a), (b) a dual-chirp LFM signal, (c), (d) a nonlinearly frequency-modulated signal, (e), (f) a step-frequency signal, and (g), (h) a frequency-hopping signal. (a)(c)(e)(g) Without feedback, and (b)(d)(f)(h) with feedback.

By using the nonlinear frequency-sweep optical signal, the time-frequency diagram of a signal can be obtained by combining the measurement results of multiple frequency-sweep periods. Moreover, when the result in a single frequency-sweep period is used, the electrical pulses after FTTM can be used to determine the frequency of the SUT in this specific period. The optoelectronic feedback can enhance the stability of the generated optical frequency-sweep signal, so the frequency measurement accuracy can also be improved when the pulses in a single period are used. Here, two two-tone signals are used for frequency measurement. The mean absolute error (MAE) of the measurement results in 80 sweep periods is used to evaluate the measurement performance. The frequency-sweep range of the pump wave is respectively set to 90, 120, 150, 180, 240, and 270 MHz. The rest of the system settings are the same as in Fig. 4. The frequency difference of the two-tone signals is set to 1 GHz and 3 GHz. Fig. 6 shows the MAE under different pump wave sweep ranges. When the sweep range of the pump wave is fixed, it is consistently observed that the MAE of the measurement results without feedback is greater than that with feedback. Besides, when the sweep range of the pump wave is set to the optimal value of 150 MHz, the MAE is minimized to smaller than 6 MHz. The consistency of the two groups of results with different frequency differences strongly verifies our conclusion in [11] that the measurement error is independent of the frequency difference. In addition, compared with the results in [11], the measurement error has been significantly reduced. There are two main reasons: 1) This work uses a feedback loop, which enables the frequency-sweep optical signal in the probe branch to have better sweeping characteristics; 2) The filter used in this work is the SBS gain spectrum, which results in pulses after FTTM with better signal-to-noise ratio.

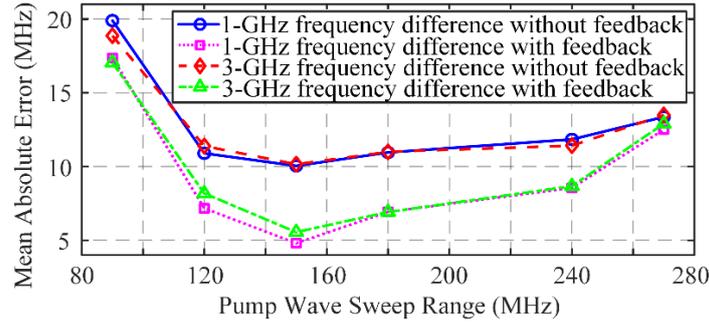

Fig. 6 Mean absolute errors of the measured frequency difference of a two-tone signal under different pump wave frequency-sweep ranges.

## 3. Conclusion

In conclusion, we have proposed and experimentally demonstrated a microwave photonic STFT system. By jointly using the optoelectronic feedback and broadened SBS gain spectrum, the time-frequency analysis resolution and frequency measurement accuracy of the method based on optical frequency-sweeping via optical injection and FTTM have been significantly improved. A method of post-compensating the influence of the nonlinearity of the optical frequency-sweep signal on the analysis result is also incorporated into the system. The method proposed in this letter is of great significance in reducing the cost of realizing STFT based on high-speed optical frequency sweeping and filtering, while also improving system performance.

**Acknowledgements**

This work was supported by the National Natural Science Foundation of China under Grant 62371191, the Shanghai Aerospace Science and Technology Innovation Foundation under Grant SAST2022-074, and the Songshan Laboratory Pre-research Project under Grant YYJC072022006.